\documentclass[prb,twocolumn,showpacs,superscriptaddress,floatfix,amsfonts,amsmath]{revtex4}
\usepackage{graphicx}

\newcommand{\be}{\begin{equation}}
\newcommand{\ee}{\end{equation}}
\newcommand{\bea}{\begin{eqnarray}}
\newcommand{\eea}{\end{eqnarray}}

\newcommand{\f}{\frac}

\newcommand{\avector}{\mathfrak{a}}
\newcommand{\dmatrix}{\mathfrak{D}}
\newcommand{\etamatrix}{\eta}
\newcommand{\bvector}{\mathfrak{b}}
\newcommand{\tmatrix}{\mathfrak{t}}
\newcommand{\ematrix}{\mathfrak{E}}

\begin{document}

\title{Spin wave calculation of the field-dependent magnetization pattern around an impurity in Heisenberg antiferromagnets} 
\author{Sergey \surname{Shinkevich}}
\affiliation{Department of Physics, University of Oslo, P.~O.~Box 1048 Blindern, N-0316 Oslo, Norway}
\author{Olav  F. \surname{Sylju{\aa}sen}}
\affiliation{Department of Physics, University of Oslo, P.~O.~Box 1048 Blindern, N-0316 Oslo, Norway}
\author{Sebastian \surname{Eggert}}
\affiliation{Department of Physics and Research Center OPTIMAS, University of Kaiserslautern, D-67663 Kaiserslautern, Germany }

\date{\today}


\begin{abstract}
We consider the magnetic-field dependent spatial magnetization pattern around a general impurity embedded in a Heisenberg antiferromagnet using both an analytical and a numerical spin wave approach. 
The results are compared to quantum Monte Carlo simulations. 
The decay of the magnetization pattern away from the impurity follows a
universal form which reflects the properties of the pure
antiferromagnetic Heisenberg model.
Only the overall magnitude of the induced magnetization depends also on 
the size of the impurity spin and the impurity coupling.
\end{abstract}

\maketitle

\section{introduction \label{introduction}}
The local magnetization around 
impurities in antiferromagnets have been studied by Nuclear Magnetic Resonance (NMR)
experiments already since the early 1970's.\cite{Butler,Wijn}. The analysis of
local Knight shifts has been expanded after the discovery of
high temperature superconductivity.\cite{alloul}  Typically, the strongly correlated state
is reflected 
by the observation of large alternating magnetic moments around
static impurities,\cite{alloul} which become especially strong in one-dimension.\cite{1d}
Another remarkable experimental tool is given by 
scanning tunneling microscopy (STM),\cite{STM} which offers 
the unique possibility of
studying materials directly on the atomic scale. In particular, by coating the STM-tips with 
different magnetic materials,\cite{SP-STM2000} 
so called spin polarized scanning tunneling microscopy (SP-STM) 
has made it possible to study the magnetization of individual atoms.\cite{SP-STMrecent}

From the theoretical point of view, antiferromagnets
 are often represented by the isotropic 
Heisenberg model with static impurities.  In this case the pinning of the order is a result 
of an interplay of the applied uniform magnetic field with impurities.
The first theoretical studies of impurities in an antiferromagnet
date back to the 1960's.\cite{Lovesey,Tonegawa}
More recent research has made much progress in the understanding of 
the impurity behavior in one-dimensional\cite{1d,affl,impsusc} 
and two-dimensional\cite{Sachdev,Sandvik,anfuso} Heisenberg antiferromagnets.
In particular, the magnetic response around a vacancy in an isotropic antiferromagnet was 
studied in Ref.~\onlinecite{PRL} using a hydrodynamic approach.
In this work, we now extend those studies by 
considering the local magnetization using spin wave theory 
for a  more general impurity type, which is given by a spin-$S_0$
coupled to the host antiferromagnet with a general coupling $J_0$.
One main result is that the decay constant of the magnetization is 
to leading order governed by properties of the host magnet, 
while the overall magnitude is governed by properties of the impurity 
and its coupling to the host antiferromagnet. We complement our 
analytical spin-wave analysis with Quantum Monte Carlo (QMC) simulations 
as well as a numerical spin wave approach for the case of 
calculating the magnetization on and close to the impurity site.    

\begin{figure}
\includegraphics[clip,width=6cm]{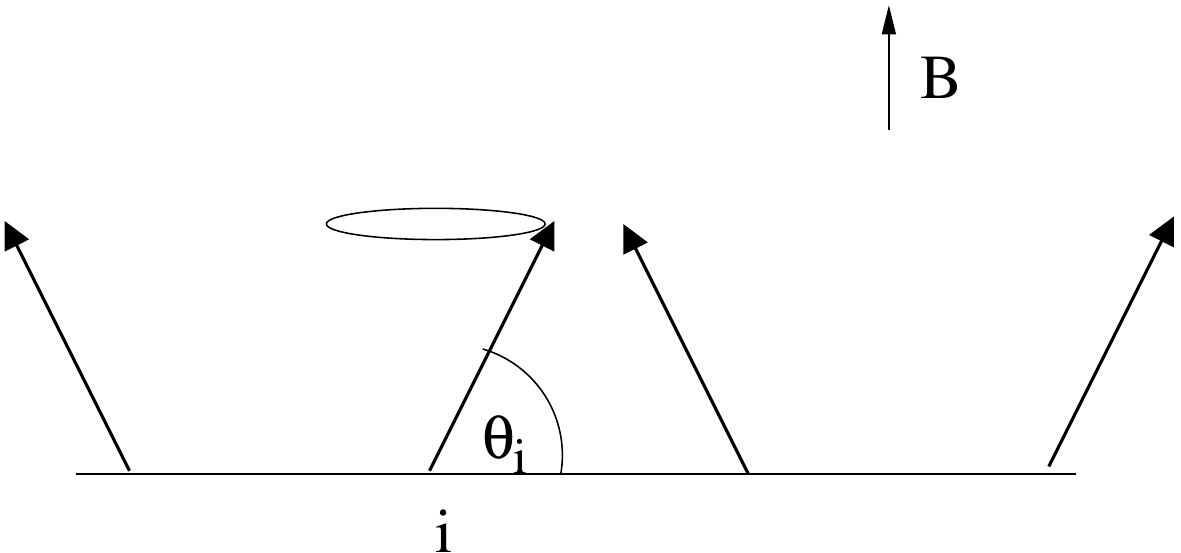} 
\caption{The canted spin state for classical spins. $\theta_i \in [0,\pi/2]$ is the angle between the spin $i$ and a line drawn perpendicular to the applied magnetic field $B$. The angle $\phi_i \in [0,2\pi]$ parametrizes how much the spin $i$ is rotated (a full rotation is indicated by the ellipse) about the applied magnetic field.\label{angles}}
\end{figure}

\section{Hamiltonian}
We consider the following Hamiltonian of a Heisenberg-type magnet in a magnetic field
\be \label{Hamiltonian}
H = \sum_{\langle i,j \rangle} J_{ij} \vec{S}_i \cdot \vec{S}_j - \sum B_i S^z_i
\ee
on a hyper cubic lattice where each site has $Z$ nearest neighbors. 
We will start out with general site dependent couplings $J_{ij}$ and magnetic fields $B_i$ and later specialize to the case of a single impurity in an otherwise uniform antiferromagnet in a homogeneous field. 

In order to treat the non-homogeneous Hamiltonian in Eq.~(\ref{Hamiltonian})
with spin wave theory, let us first review in detail how to derive 
the expansion in fluctuations about an ordered classical state.
The classical state of an antiferromagnet in a magnetic field is that of canted spins pointing partly along the z-axis, see Fig.~\ref{angles}. In order to parametrize this state we introduce rotated spins $\vec{S}^\prime$ so that $S_i^{\prime z}$ points along a direction parametrized by the angles $\theta_i$ and $\phi_i$, see Fig.~\ref{angles}. 

The rotated spin components $\vec{S}^\prime$ are related to the spin components in Eq.~(\ref{Hamiltonian}) as
\bea 
S^{x}_i & = & \left( S^{\prime x}_i \sin \theta_i - S^{\prime z}_i \cos \theta_i \right) \cos \phi_i - S^{\prime y}_i \sin \phi_i \nonumber \\
S^{y}_i & = & \left( S^{\prime x}_i \sin \theta_i - S^{\prime z}_i \cos \theta_i \right) \sin \phi_i - S^{\prime y}_i \cos \phi_i  \label{rotation} \\
S^{z}_i & = & S^{\prime x}_i \cos \theta_i + S^{\prime z}_i \sin \theta_i. \nonumber
\eea
Inserting these into Eq.~(\ref{Hamiltonian}) we get the Hamiltonian expressed 
in terms of rotated spins for arbitrary angles, which will be determined later.   
In order to express 
the fluctuations about the ordered state we use the Holstein-Primakoff 
transformation\cite{HolsteinPrimakoff} on the rotated spins
into bosonic operators 
\bea
S^{\prime z}_i & = & S_i - a^\dagger_i a_i \nonumber \\
S^{\prime +}_i & = & \sqrt{2S_i}  \sqrt{1 - \f{a^\dagger_i a_i}{2s}} \, a_i \\
S^{\prime -}_i & = & \sqrt{2S_i} \, a^\dagger_i \sqrt{1 - \f{a^\dagger_i a_i}{2s}} \nonumber
\eea
where
expanding the square roots and using $S^{\prime \pm}_i = S^{\prime x} \pm i S^{\prime y}$ yields
\bea
S^{\prime x}_i & = & \sqrt{\f{S_i}{2}} \left( a_i + a^\dagger_i - \f{1}{4s} \left( a^\dagger_i a_i a_i+a^\dagger_i a^\dagger_i a_i \right)+\ldots \right) \\
S^{\prime y}_i & = & -i \sqrt{\f{S_i}{2}} \left( a_i - a^\dagger_i - \f{1}{4s} \left( a^\dagger_i a_i a_i-a^\dagger_i a^\dagger_i a_i \right)+\ldots \right). \nonumber
\eea
By inserting these expressions for $\vec{S}^{\prime}$ into the Hamiltonian Eq.~(\ref{Hamiltonian}) we get terms $H_n$ with different powers $n$ of bosonic operators. 

\begin{figure}
\includegraphics[clip,width=3.5cm]{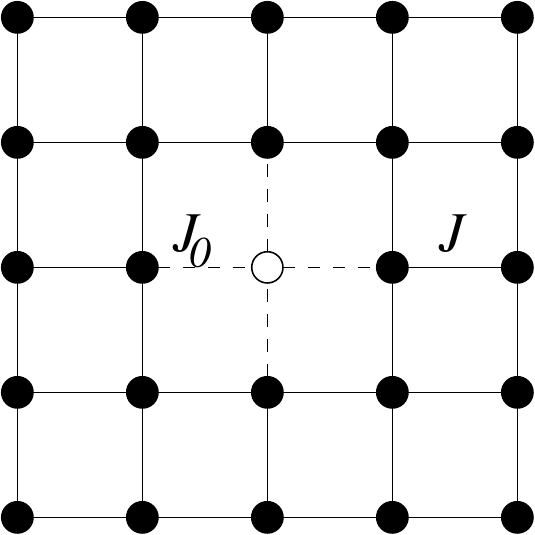} 
\caption{Couplings. Dashed lines indicate the coupling $J_0$ to the impurity site (empty circle) while solid lines indicate $J$. \label{couplings}}
\end{figure}

The zeroth order term in boson operators corresponds to the energy of classical spins oriented along the $S^{\prime z}$ axes. This is so because in the classical limit $S_i \to \infty$ the $S^{\prime x}$ and $S^{\prime y}$ components are overwhelmed by the $S^{\prime z}$ component which is proportional to $S$. The zeroth order terms read
\bea
H_0 \! = & & \!\!\!\!\!\!    \sum_{\langle ij \rangle} J_{ij} S_i S_j \left( \cos \theta_i \cos \theta_j \cos( \phi_{ij}) + \sin \theta_i \sin \theta_j \right) \nonumber \\
& &     - \sum_i B_i S_i \sin \theta_i    \label{zeroth}
\eea
where $\phi_{ij} = \phi_i - \phi_j$.
Because of the U(1) symmetry of spin rotations about the 
magnetic field axis $H_0$ depends on the 
relative angles $\phi_{ij}$. 
Minimizing with respect to $\phi_{ij}$ gives the condition
\be
- J_{ij} S_i S_j \cos \theta_i \cos \theta_j \sin( \phi_{ij} ) = 0
\ee
meaning that $\phi_{ij} = 0$ or $\pi$. For this to be a {\em minimum} of the energy one needs $-J_{ij} \cos( \phi_{ij} ) > 0$, which means that $\phi_{ij}=\pi$ for an antiferromagnetic coupling and 0 for a ferromagnetic one. Equivalently $-\cos( \phi_{ij} ) = J_{ij}/|J_{ij}| \equiv \nu_{ij}$.
We will in the following select the rotation angle $\phi_0$ so that it is either $0$ or $\pi$. With this choice, and the minimization condition $\phi_{ij}=0$ or $\pi$, all terms with $\sin \phi_i$ will be zero. Then the Hamiltonian can be written
\bea
H & = &\sum_{<ij>} J_{ij} \left[ \cos(\theta_i + \nu_{ij} \theta_j) \left( S^{\prime x}_i S^{\prime x}_j 
						  - \nu_{ij}S^{\prime z}_i S^{\prime z}_j \right) \right.\nonumber \\
& &                \left.
	      -\nu_{ij} S^{\prime y}_i S^{\prime y}_j 
	      +\sin(\theta_i+\nu_{ij} \theta_j) 
	      \left( \nu_{ij} S^{\prime x}_i S^{\prime z}_j + S^{\prime z}_i S^{\prime x}_j \right) \right] \nonumber \\
& & -\sum_i B_i \left(  S^{\prime x}_i \cos \theta_i + S^{\prime z}_i \sin \theta_i \right).
\label{Ham2}
\eea
We will now specialize to the case of a single impurity embedded in an otherwise uniform antiferromagnet of spin-$S$ spins. We label the impurity site $i=0$ and allow for an impurity spin $S_0$ which in general can be different from $S$. We take all bonds not connected to the impurity to be antiferromagnetic with a magnitude $J$. The bonds connected to the impurity are also equal, but of a different magnitude $J_0$ and can be either ferromagnetic or antiferromagnetic, see Fig.~\ref{couplings}, $\nu_0$ denotes the sign of $J_0$. This antiferromagnet is placed in a magnetic field oriented along the $z$-direction with magnitude $B$. We have absorbed the Zeeman coupling into the magnitude of the magnetic field. In order to allow for a different gyromagnetic factor of the impurity spin and thus a different Zeeman coupling, we label the magnitude of the effective magnetic field on the impurity site $B_0$ which in general can be different from $B$. 

In order to simplify Eq.~(\ref{Ham2}) we use 
an initial rotated frame that is given by 
a site independent value of $\theta_i= \theta$ 
for all sites $i$ away from the impurity site to zeroth order.
We will later allow 
for a site-dependent shift of $\theta$ in order to calculate the 
non-trivial local variation of the magnetization.
For the impurity site $i=0$ we keep a separate angle $\theta_0$. 
Performing this ansatz the zeroth order term in boson operators takes the form
\bea
H_0 & = & - NS \left( \f{JSZ}{2} \cos 2\theta + B \sin \theta \right) \nonumber \\
&   & + ZS\left( \vphantom{\f{1}{2}} JS \cos 2\theta - |J_0| S_0 \cos(\theta+\nu_0 \theta_0) \right) \\
&   & + BS \sin \theta - B_0 S_0 \sin \theta_0. \nonumber 
\eea
Minimizing this with respect to $\theta$ and $\theta_0$ in the thermodynamic limit, $N \to \infty$, determines the angles $\theta$ and $\theta_0$
\be \label{uniformcondition}
\sin \theta = \f{B}{2SZ J}
\ee
and
\be \label{impuritycondition}
\tan \theta_0 = \f{B_0}{|J_0| S Z \cos \theta} -\nu_0 \tan \theta.
\ee
The zeroth order 
condition on $\theta$ is identical to the one found for a uniform antiferromagnet 
in a homogeneous field and does not depend on the impurity. This is a natural consequence 
of taking a site-independent ansatz in the thermodynamic limit.

When using the value of $\theta$ obtained from Eq.~(\ref{uniformcondition}) 
the terms that are of linear order in boson operators
 connected to the bulk behavior vanish. 
After also using the condition Eq.~(\ref{impuritycondition}) only 
linear
terms of bosons around the impurity are left
\be
H_1 = \f{C}{Z} \sum_{<0j>} \left( a_{j} + a^\dagger_{j} \right)
\ee
where the sum is restricted to run over the nearest neighbors of the impurity spin. 
This expression can be interpreted as a local effective field in the rotated frame 
acting on the 
spins that are coupled to the impurity spin, which will cause a shift of the 
angles $\theta$ over an extended range as we will see later.

The constant $C$ is given by
\be 
C =  J_0 S_0 Z \sqrt{\f{S}{2}} 
\nu_0 \sin(\nu_0 \theta_0+\theta) - JSZ \sqrt{\f{S}{2}} \sin 2\theta
\ee
or equivalently 
when we use the minimization conditions
\be \label{eqforC}
C= \sqrt{\f{S}{2}} \left( \f{S_0}{S} \nu_0 B_0 \cos \theta_0 - B \cos \theta \right).
\ee
The linear terms can also be written in terms of Fourier transforms 
\be
a_i = \f{1}{\sqrt{N}} \sum_{\vec{k}} a_{\vec{k}} e^{i\vec{k} \cdot \vec{r}_i}
\ee
as
\be \label{linearterm}
H_1 = \f{C}{\sqrt{N}} \sum_{\vec{k}} \gamma_{\vec{k}} \left( a_{\vec{k}} + a^\dagger_{\vec{k}} \right)
\ee
where we have defined $\gamma_k = 2(\cos k_x + \cos k_y + \ldots)/Z$ where the $k$'s are given in units of the inverse lattice spacing. 

For the quadratic terms we will as a first approximation keep only the terms that are
leading order in $N$. 
Therefore, the quadratic terms are identical to those in the absence of an impurity 
\be \label{bulkH}
H_{2}^{\rm bulk} = \f{1}{2} \sum_{\vec{k}} \left\{ A_{\vec{k}} a^\dagger_{\vec{k}} a_{\vec{k}} + B_{\vec{k}} a_{\vec{k}} a_{-\vec{k}} + h.c. \right\}
\ee
where $A_{\vec{k}} = JSZ( \cos 2\theta -\gamma_{\vec{k}} \sin^2 \theta) + B\sin \theta
 = JSZ(1-\gamma_{\vec{k}} \sin^2 \theta)$ 
and $B_{\vec{k}} =JSZ \cos^2 \theta \gamma_{\vec{k}}$ 
which are also known from standard spin-wave theory.\cite{ZhitomirskyChernyshev} 
The neglected quadratic impurity terms can in principle
lead to a renormalization of the overall magnitude in the local order 
around the impurity.  However, this effect is known to be surprisingly small
from numerical studies,\cite{bulut}
so that we can omit those terms for now in order to calculate
the magnetization around the impurity.  We will include them later when considering
the magnetization of the impurity spin itself.

The quadratic term can be diagonalized by the canonical transformation
\be
a_{\vec{k}} = u_{\vec{k}} b_{\vec{k}} + v_{\vec{k}} b^\dagger_{-\vec{k}}
\ee
which results in the quadratic Hamiltonian
\be \label{bulkhamiltonian}
H_{2}^{\rm bulk} = \sum_{\vec{k}} \omega_{\vec{k}} b^\dagger_{\vec{k}} b_{\vec{k}} + \f{1}{2} \sum_{\vec{k}} \left( \omega_{\vec{k}} - A_{\vec{k}} \right)
\ee	  
where $\omega_{\vec{k}} = \sqrt{A_{\vec{k}}^2 - B_{\vec{k}}^2}$ which becomes  
\be \label{dispersion}
\omega_{\vec{k}} = JSZ \sqrt{ \left( 1-\gamma_{\vec{k}} \right) \left( 1+\cos 2\theta \gamma_{\vec{k}} \right)}.
\ee 
The transformation coefficients obey $u_{\vec{k}}^2-v_{\vec{k}}^2=1$, $u_{\vec{k}}^2+v_{\vec{k}}^2=A_{\vec{k}}/\omega_{\vec{k}}$ and $2u_{\vec{k}} v_{\vec{k}} = -B_{\vec{k}}/\omega_{\vec{k}}$.

Using the quadratic bulk Hamiltonian we
can calculate the following expectation values
\bea
\delta & = & \langle a_i a_i \rangle = \f{1}{N} \sum_{\vec{k}} u_{\vec{k}} v_{\vec{k}} \nonumber \\
\Delta & = & \langle a_i a_j \rangle = \f{1}{N} \sum_{\vec{k}} \gamma_{\vec{k}} u_{\vec{k}} v_{\vec{k}} \nonumber \\
m & = & \langle a^\dagger_i a_j \rangle = \f{1}{N} \sum_{\vec{k}} \gamma_{\vec{k}} v_{\vec{k}}^2 \nonumber \\
n & = & \langle a^\dagger _i a_i \rangle = \f{1}{N} \sum_{\vec{k}} v_{\vec{k}}^2 \label{n}
\eea
for nearest neighbor sites $i$ and $j$. Note that the bulk nature of the quadratic term dictates
that these expressions do not depend on $i$ and $j$. At this stage we truncate higher order terms in the Hamiltonian.  Therefore we have reduced the problem to a solvable bulk Hamiltonian
in Eq.~(\ref{bulkH}) together with an impurity term in Eq.~(\ref{linearterm}).

\section{Magnetization away from the impurity}
The magnetization in the direction of the field $M^z_i = \langle S^z_i \rangle$ is
\be
M^z_i = \langle S^{\prime x}_i \rangle \cos \theta_i + \langle S^{\prime z}_i \rangle \sin \theta_i.
\ee
Expressed in terms of bosons the above expression is up to quadratic order
\be
M^z_i \approx \sin \theta_i \left( S_i - \langle a^\dagger_i a_i \rangle \right) + \cos \theta_i \sqrt{\f{S_i}{2}} \left( \langle a^\dagger_i \rangle+\langle a_i \rangle \right).
\ee
To calculate these expectation values in the presence of the impurity
we perform a shift of the boson operators
\be
a_i \to a_i + \alpha_i
\ee
so as to get rid of the remaining linear terms in the Hamiltonian in Eq.~(\ref{linearterm}).
 This is equivalent to 
a site dependent variation of the angle $\theta_i$.
The impurity induced shift is given by 
\be \label{alpha}
\alpha_i =  -\f{C}{N} \sum_{\vec{k}} \f{\gamma_{\vec{k}}}{A_{\vec{k}}+B_{\vec{k}}} e^{i \vec{k} \cdot \vec{r}_i}.
\ee
For future convenience we parametrize
\be
A_{\vec{k}}+B_{\vec{k}} = f\left(1 + g \gamma_{\vec{k}} \right)
\ee
in terms of constants $f$ and $g$ which to leading order in $1/S$ are obtained from Eq.~(\ref{bulkH}); $f=JSZ$ and $g=\cos 2\theta$.

Shifting the boson operators gives the following expression for the magnetization
\be
M^z_i \approx \sin \theta_i \left( S_i - |\alpha_i|^2 - \langle a^\dagger_i a_i \rangle \right)+ \sqrt{\f{S_i}{2}} \cos \theta_i \left( \alpha^*_i +\alpha_i \right).
\ee
Since the shift of the boson operators has eliminated the linear terms, we can now 
use the usual bulk theory to calculate the corresponding expectation value
$n= \langle a^\dagger_i a_i \rangle$ in Eq.~(\ref{n}).
Thus the magnetization takes the form
\be
M^z_i \approx \sin \theta \left( S - |\alpha_i|^2 - n \right)+ \sqrt{\f{S}{2}} \cos \theta \left( \alpha^*_i +\alpha_i \right) , i \neq 0.
\ee
As is shown in the Appendix, $\alpha_i$ is real and changes sign depending on 
which sublattice $i$ belongs to with 
$e^{i \vec{Q} \cdot \vec{r}} = (-1)^{x_i+y_i+z_i}$ where 
$\vec{Q}=(\pi,\pi,\ldots)$ 
is the antiferromagnetic 
wave vector.
Therefore, it is convenient to write 
$\alpha_i = (-1)^{x_i+y_i+z_i} \tilde{\alpha}_i$ and to 
divide the magnetization into an alternating and a non-alternating part. 
Using the assumption that $\tilde{\alpha}_i$ does not vary rapidly, 
the alternating(non-alternating) magnetization on site $i$ is obtained by taking half of the magnetization on an odd sublattice site $i$ and subtract (add) half of the magnetization on the neighboring even sublattice sites surrounding site $i$.  
Therefore, the non-alternating part takes the form
\be
M^z_{{\rm nalt},i} = \sin \theta \left( S-n-{\tilde{\alpha}}_i^2 \right)
\ee
which will decay rapidly to its uniform bulk value. This non-alternating part is not our primary focus here. Instead we will focus on the alternating part which does not decay as rapidly. To leading order the alternating magnetization is
\be \label{maltdef}
M^z_{{\rm alt},i} = -\sqrt{2S} \cos \theta \; \tilde{\alpha}_i,
\ee
thus $\tilde{\alpha}_i$ dictates its behavior. 
The sum in Eq.~(\ref{alpha}) can be carried out by expanding the integrand about the minimum of the denominator which is at the antiferromagnetic point $\vec{Q}=(\pi,\pi,\ldots)$ as shown in the Appendix. 
Carrying out this expansion for the case $i \neq 0$, we get in $D=2$ and $D=3$ dimensions
\be \label{tildealpha}
\tilde{\alpha}_i \approx \f{C Z}{2\pi f g^2}    
\left\{ 
\begin{matrix} 
K_0(r_i/d),  & \; D=2 \\
~& ~\\
e^{-r_i/d}/(2r_i),& \; D=3
\end{matrix} \right. \; , i \neq 0.  
\ee
where $r_i=\sqrt{x_i^2+y_i^2+z_i^2}$ is the distance from the impurity in units of the lattice spacing and $K_0$ is the zeroth order modified Bessel function of the second kind which decays as $e^{-r_i/d}/\sqrt{r_i}$ for large arguments. The characteristic decay scale is 
\be \label{decay}
d = \sqrt{ \f{g}{Z(1-g)}}
\ee
in both cases.  The result in Eq.~(\ref{tildealpha}) is the main result of this section for
the induced magnetization by the general impurity model, which will be compared to 
Monte Carlo results in the following.
Note, that the shape and the decay scale $d$ is universal and 
only depends on properties of the host magnet in the bulk.
Only the constant prefactor $C$ in Eq.~(\ref{eqforC}) depends on 
impurity properties $S_0$, $J_0$ and $B_0$. 
With the expression $g=\cos 2\theta$, the decay constant is 
$d=[\cos 2\theta/(2Z \sin^2 \theta)]^{1/2}$. 

\begin{figure}
\includegraphics[clip,width=8cm]{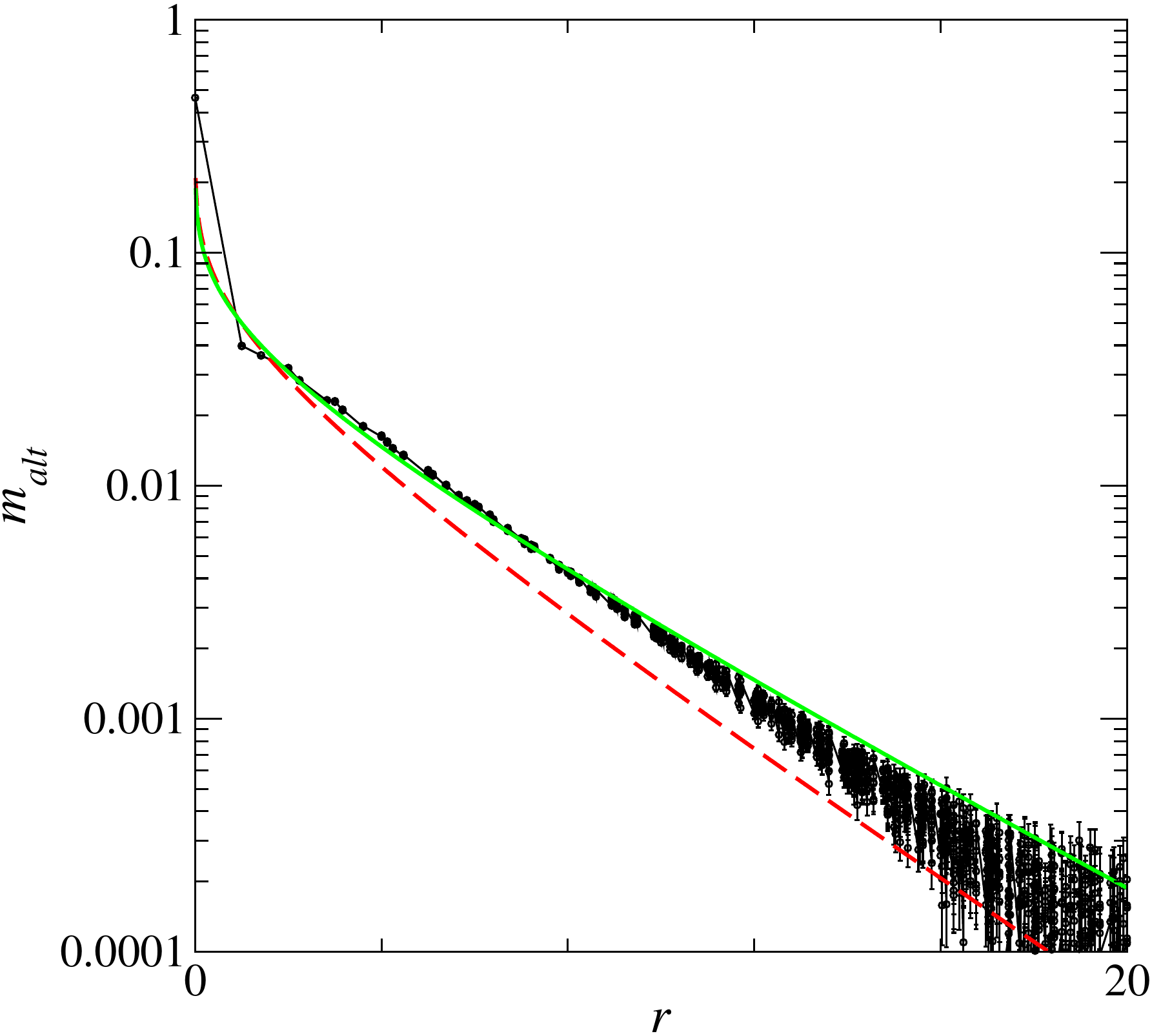} 
\caption{(color online) $M^z_{alt}$ vs. distance from the impurity $r$ on the square lattice. The circles are quantum Monte Carlo data while the dashed line (red) is a plot of the 
analytic result Eq.~(\ref{maltdef}) using $g=\cos 2\theta$. The result where we have taken into account $1/S$-corrections for $A_{\vec{k}}+B_{\vec{k}}$ is shown as the solid line (green). 
Here $S=S_0=1/2$, $Z=4$, $B=B_0=0.4J$ and $J_0=0.1J$. 
\label{maltdecay}} 
\end{figure}

In Fig.~\ref{maltdecay} we have plotted a comparison of $M^z_{\rm alt}$ calculated using the 
expression in Eqs.~(\ref{maltdef})-(\ref{tildealpha}) and results from a QMC simulation. The QMC simulations were carried out using the stochastic series expansion technique\cite{SSE} using directed-loop updates\cite{SS} at a low temperature $T/J=0.05$ on a $128 \times 128$ square lattice. As can be seen from Fig.~\ref{maltdecay} the leading order analytical result decays faster than the QMC result. However the decay $d$ depends crucially on the exact expression for $A_{\vec{k}}+B_{\vec{k}}$ which we have approximated with its leading order value $d=[\cos 2\theta/(2Z \sin^2 \theta)]^{1/2}$. In fact, we can do better by including $1/S$ corrections. Taking into account $1/S$ corrections to $A_{\vec{k}}+B_{\vec{k}}$ and to the angle $\sin \theta$, we get
\bea
A_{\vec{k}}+B_{\vec{k}} & = & JSZ \left[ 1- \frac{2n+2\Delta+m}{2s}-\sin^2 \theta \frac{m+\Delta}{2s} \right. \nonumber \\
&   &+ \gamma_{\vec{k}} \left( \cos{2\theta}-\frac{2n+2m+2\Delta+\delta}{2s} \right. \nonumber \\
&   &\left. \left. -\sin^2{\theta} \frac{2n+2m+2\Delta-\delta}{s} \right) \right].
\eea
This result can also be inferred from Ref.~\onlinecite{ZhitomirskyChernyshev}.
The $1/S$ corrections give modified expressions for the constants $f$ and $g$, which 
leads to a better agreement with the Monte Carlo data in Fig.~\ref{maltdecay}.
By allowing also another classical angle $\theta_1$ for the impurity nearest neighbor spins the agreement with QMC close to the impurity site can be improved at the expense of having more complicated analytic expressions.  
\begin{figure}
\includegraphics[clip,width=8cm]{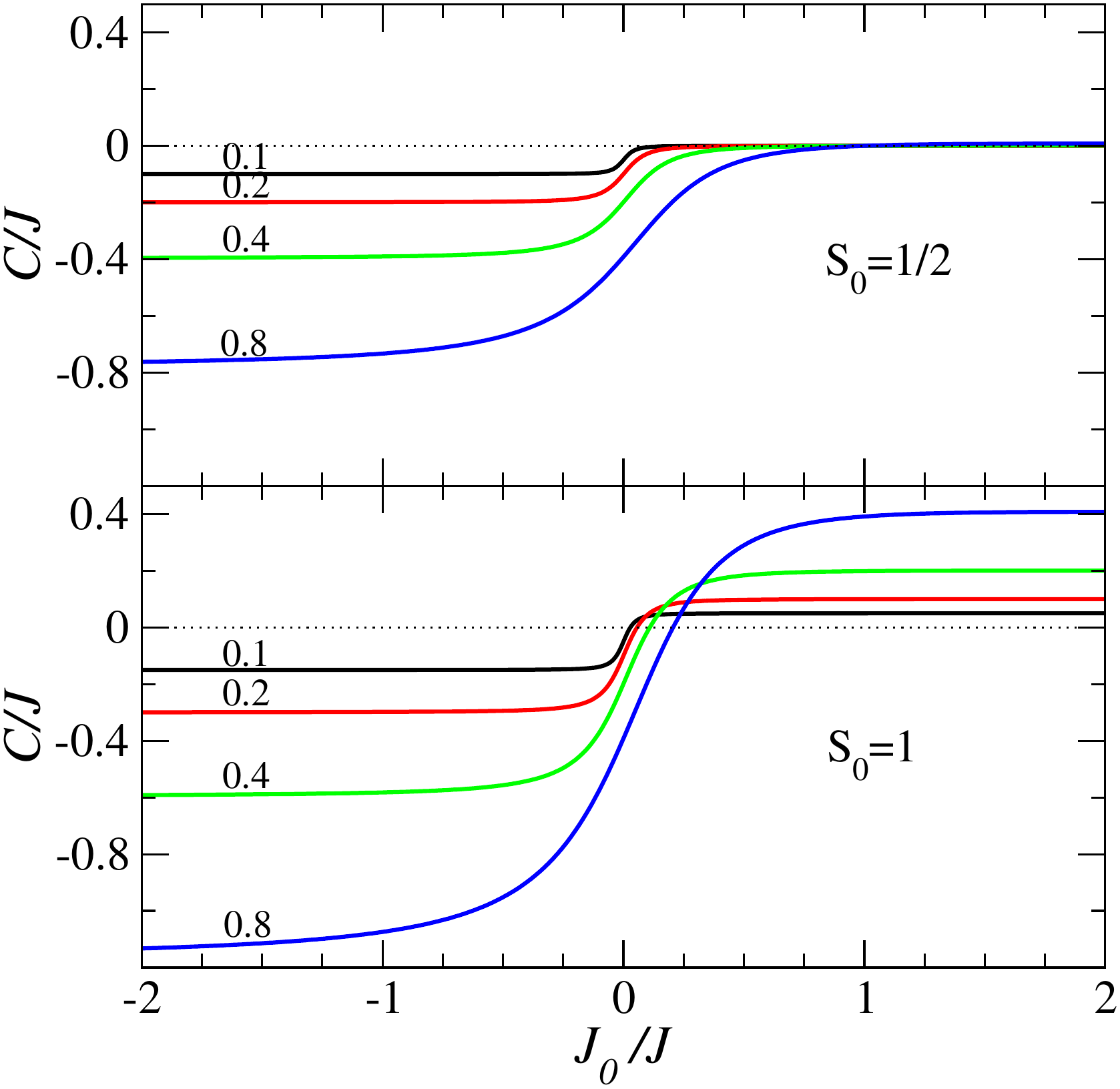} 
\caption{(color online) $C/J$ vs. impurity coupling $J_0$ for impurity spin $S_0=1/2$ (upper panel) and $S_0=1$ (lower panel) for different values of the magnetic field $B/J$ indicated by the numbers above each curve on the left side. Here $S=1/2$ and $Z=4$, $B_0=B$. 
\label{C}} 
\end{figure}
To connect our result in Eqs.~(\ref{maltdef})-(\ref{tildealpha}) to that obtained in Ref.~\onlinecite{PRL} for the induced magnetization around a vacancy ($J_0=0$) we observe that for $\vec{k}$ close to $\vec{Q}$ but $|\vec{k}-\vec{Q}| > [8 \sin^2 \theta/\cos 2\theta]^{1/2}$ the dispersion Eq.~(\ref{dispersion}) is linear with a spin-wave velocity $c=2JS \sqrt{2\cos 2\theta}$. In the limit $B \to 0$ this becomes the well-known leading order spin wave theory result for the spin wave velocity of an antiferromagnet. Combining this with Eq.~(\ref{uniformcondition}) we see that the decay constant of Ref.~\onlinecite{PRL} becomes
 $c/B = [\cos 2\theta/(8 \sin^2 \theta)]^{1/2}$,
which equals the leading order result for the decay constant $d$.
Similarly, we can compare the factor multiplying the Bessel-function $K_0$. In the case of a vacancy $J_0=0$ our expression for $C=-(S/2)^{1/2} B \cos \theta$ so that the prefactor becomes
\be
-\sqrt{2S} \cos \theta \f{C}{2\pi f g^2} \approx \f{B}{2\pi J}
\ee
where we have used $f=JSZ$ and $g=\cos 2 \theta$ and approximated $\cos \theta \approx 1$ which is valid for low magnetic fields. This is to be compared to the expression $m_{\rm max} S B/(2\pi \rho_s)$ obtained in Ref.~\onlinecite{PRL}. When inserting the leading order expression $m_{\max}=S$, $\rho_s=JS^2$ we see that the two results become equal.

\begin{figure}
\includegraphics[clip,width=8cm]{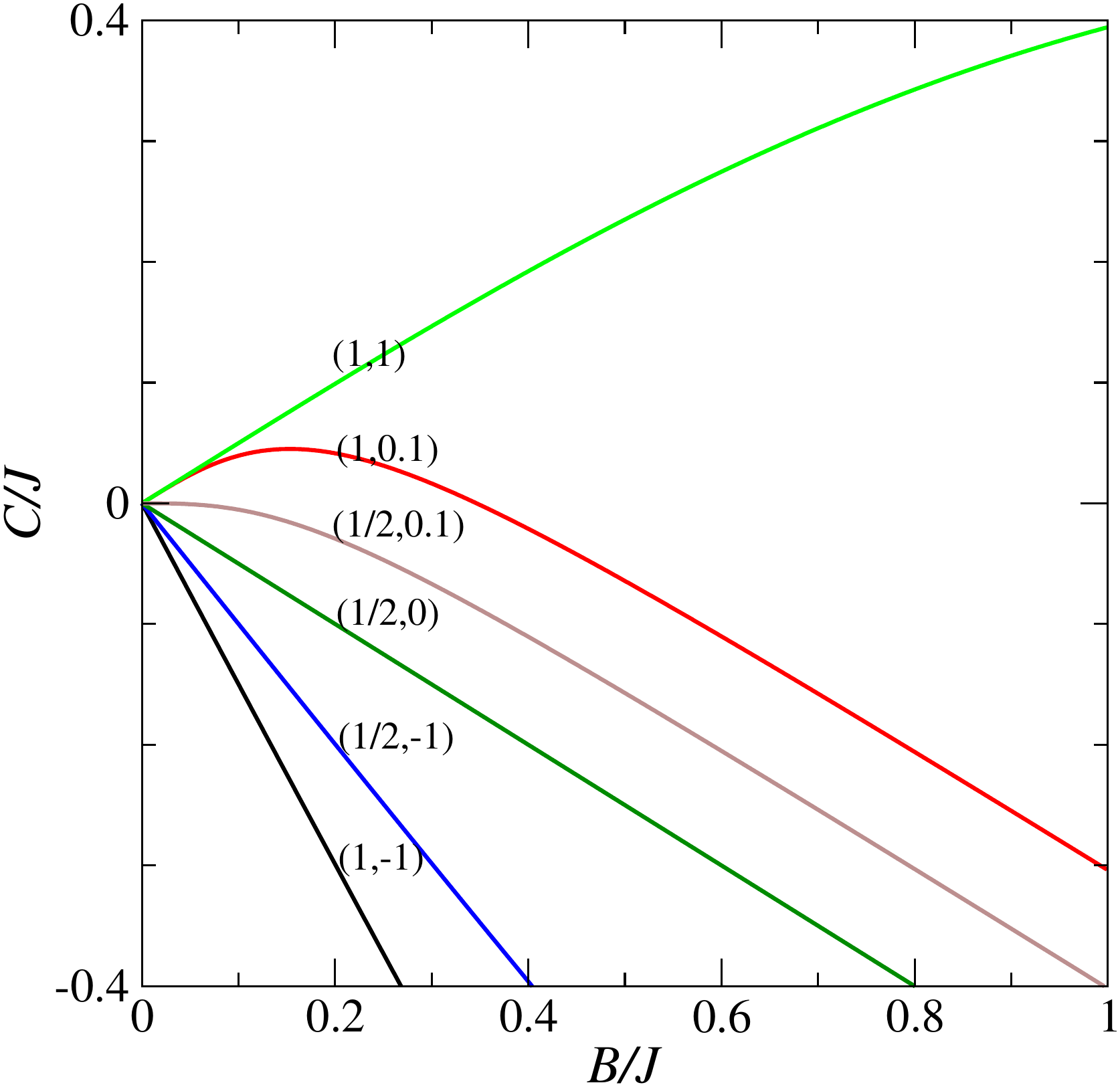} 
\caption{(color online) $C/J$ vs. magnetic field $B/J$ for different values of the impurity spin and coupling denoted by $(S_0,J_0)$. Here $B_0=B$, $S=1/2$ and $Z=4$.  
\label{CvsB}} 
\end{figure}
For larger fields the use of the renormalized zero field spin-wave velocity $c$ 
in Ref.\onlinecite{PRL} is not so natural, however. 
As the decay depends heavily on the behavior of $A_{\vec{k}}+B_{\vec{k}}$ around $\vec{k}=\vec{Q}$ where the dispersion is quadratic in a finite field,
it is more natural to relate the decay constant to the effective mass of this minimum. For finite but not too large fields the dispersion around $\vec{Q}$ can be written $\omega_{\vec{k}} = B+\f{\vec{k}^2}{2m}$ where the effective mass is $m=\f{2Z \sin^2 \theta}{B \cos 2\theta}$. It is then straightforward to see that the leading order decay constant also can be written $d=1/\sqrt{Bm}$. 

\begin{figure}
\includegraphics[clip,width=6cm]{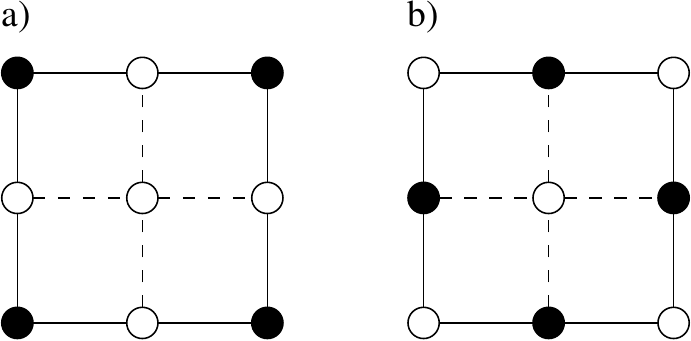} 
\caption{Orientations of the magnetization close to the impurity. The impurity spin is the middle circle. Open circles indicate that the magnetization is pointing along the applied magnetic field while filled circles indicate the opposite orientation. a) $C<0$.  b) $C>0$.   
\label{spins}} 
\end{figure}

While the decay of the induced alternating magnetization pattern is governed by the properties of the uniform magnet, the {\em magnitude} of the alternating magnetization 
is given in terms of the prefactor $C$ in 
Eq.~(\ref{eqforC}), which depends on impurity properties as shown in Figs.~\ref{C} and \ref{CvsB}. For impurity spin $S_0=1/2$ and coupling  $0< J_0<1$, the prefactor $C$ is negative and rather small.
For $J_0=J$ it vanishes completely because it corresponds to the uniform case. For ferromagnetic couplings $J_0<0$, $|C|$ gets larger with increasing magnetic field $B/J$.
Thus we expect a substantial induced alternating magnetization pattern for 
ferromagnetically coupled impurities. 
Note, however, that when the field gets larger the magnetization pattern decays faster 
with distance from the impurity.  For an $S_0=1$
 impurity, $|C|$ is no longer necessarily small for antiferromagnetic couplings and it changes sign at a small positive value of $J_0/J$. The sign change signals a sublattice change  in the magnetization pattern as indicated in Fig.~\ref{spins}, where for a ferromagnetic impurity the magnetization follows the pattern shown in Fig.~\ref{spins}~a). This pattern extends also to weak antiferromagnetic couplings up to a critical value of $J_0$ that depends on the magnetic field where it becomes favorable to interchange the orientation of magnetization on the two sublattices while keeping the impurity spin oriented along the field. This results in the pattern shown in Fig.~\ref{spins} b).  
For large values of $B/J$ and for all couplings except large antiferromagnetic ones,
 $|C|$ increases linearly with field strength $B/J$ as shown in Fig.~\ref{CvsB}. For $S_0=1$ and a small antiferromagnetic coupling $J_0$, $C$ changes sign as the magnetic field is increased, second curve from the top in Fig.~\ref{CvsB}. Thus a change in the 
sublattice rearrangement in Fig.~\ref{spins} can also happen for a fixed $J_0$ as the magnetic field is varied.
The exact point where $C$ reverses sign is special, because when $C=0$ the spin-1 impurity appears to have no effect on the host spins of the surrounding antiferromagnet.  Therefore, the field and/or the coupling can be tuned in such a way that the impurity becomes almost invisible to the bulk, i.e. very little scattering occurs.

\section{Magnetization of the Impurity Spin} \label{impmag}
\begin{figure}
\includegraphics[clip,width=8cm]{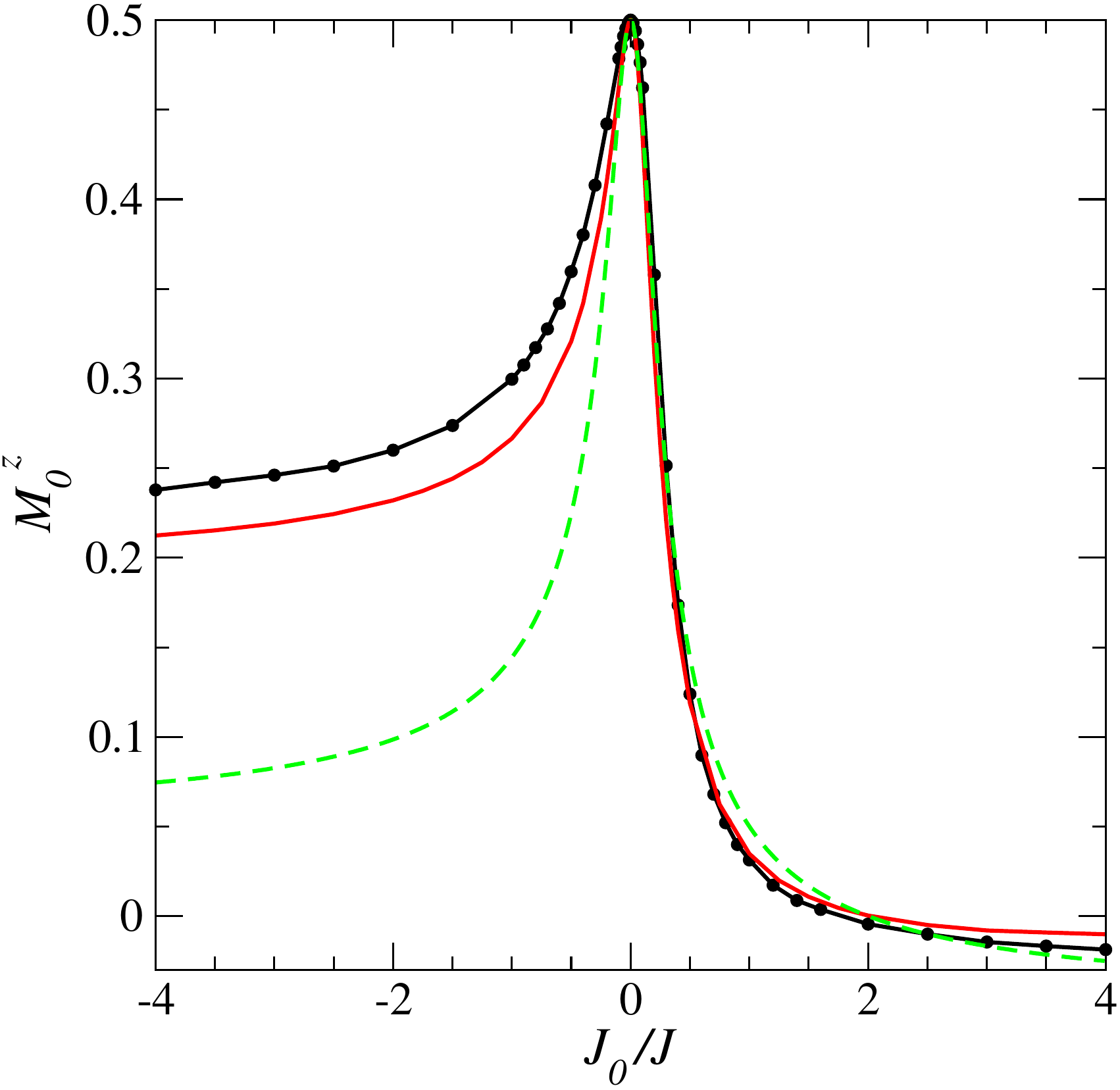} 
\caption{(color online) Magnetization at the impurity site for a spin-1/2 impurity coupled to a bulk spin-1/2 antiferromagnet by a coupling $J_0$. The filled black circles are results from Quantum Monte Carlo simulations. The dashed line (green) is the classical result coming from Eq.~(\ref{classicalmag}), and the solid line (red) is the numerical spin wave result.  
\label{impmagnS12}} 
\end{figure}

At the impurity site the leading order magnetization is obtained by the classical expression
\be
M_0^z = S_0 \sin \theta_0. \label{classicalmag}
\ee
For $S_0=1/2$ and $J_0>0$ this gives a reasonable agreement with the QMC data, 
as is seen in Fig.~\ref{impmagnS12}. 
However for other spins and ferromagnetic couplings $J_0<0$ the result 
is rather far of the QMC result. 
Thus it is necessary to also take into account the 
quantum corrections to Eq.~(\ref{classicalmag}). 
However, these quantum corrections are difficult to calculate analytically. 
This is because for the impurity itself it is necessary to include explicitly the bilinear terms connecting the impurity site to its neighbors
in addition to the quadratic bulk part in Eq.~(\ref{bulkhamiltonian}). 
These impurity terms induce non-local interactions in $k$-space, 
thus an analytic diagonalization becomes difficult. 
In order to solve this we will instead 
numerically diagonalize the quadratic boson Hamiltonian as
described below, which gives much better results shown in Fig.~\ref{impmagnS12}. 
As this method is numerical there is no need for the restriction of keeping only two angles $\theta_0$ and $\theta$. Thus we will instead keep track of all the angles $\theta_i$. 
This has the consequence 
that all linear boson terms vanish 
when using the values of the angles obtained from minimizing the zeroth order term, 
as will be shown below. 

As a function of all angles $\theta_i$ 
the zeroth order term is 
\be
H_0 = \sum_{<ij>} -|J_{ij}|S_i S_j \cos( \theta_i + \nu_{ij} \theta_j ) - \sum_i B_i S_i \sin \theta_i
\ee
where we have used the minimization condition for the $\phi$'s.
Minimizing $H_0$ with respect to $\theta_i$ we find
\be \label{minimalcondition}
\sum_{j=e_i} | J_{ij} | S_j \sin( \theta_i + \nu_{ij} \theta_j) - B_i \cos \theta_i =0 
\ee
where the sum is restricted to run over the nearest neighbors $e_i$ of site $i$. This condition is
equivalent to the equation
\be \label{allminimal}
\tan \theta_i = \f{ B_i - \sum_{j=e_i} J_{ij} S_j \sin \theta_j}{\sum_{j=e_i} |J_{ij}| S_j \cos \theta_j}.
\ee
The operators $S^{\prime x}_i S^{\prime z}_j$, $S^{\prime z}_i S^{\prime x}_j$ and the magnetic field term in Eq.~(\ref{Ham2}) give the linear terms of the Hamiltonian
\bea
H_1 & = &\sum_{<ij>} 
\left( |J_{ij}| \sqrt{\f{S_i}{2}} S_j \sin( \theta_i+\nu_{ij} \theta_j) \left( a_i+a_i^\dagger \right)
\right. \nonumber \\
&  &\left. \vphantom{\sqrt{\f{S_i}{2}}}+(i \leftrightarrow j) \right)- \sum_i B_i \sqrt{\f{S_i}{2}} \cos \theta_i \left( a_i + a_i^\dagger \right) \nonumber \\
& = & \sum_i \sqrt{\f{S_i}{2}} \left( a_i + a_i^\dagger \right) \times \nonumber \\
&   & \left( \sum_{j=e_i} | J_{ij}| S_j \sin( \theta_i+\nu_{ij} \theta_j) - B_i \cos \theta_i \right).
\eea
By comparing this to Eq.~(\ref{minimalcondition}) we see that the minimization of the constant terms leads to the vanishing of the linear terms.

The quadratic terms are
\bea
H_2 & = &\sum_{<ij>} 
J_{ij} \sqrt{\f{S_i S_j}{4}} \left( \cos(\theta_i +\nu_{ij} \theta_j)-\nu_{ij} \right) 
\left(a^\dagger_i a_j +a^\dagger_j a_i \right) \nonumber \\
&  &
+J_{ij} \nu_{ij} \cos(\theta_i +\nu_{ij} \theta_j) \left( S_j a^\dagger_i a_i +S_i a^\dagger_j a_j \right) 
\nonumber \\
&  &
+J_{ij} \sqrt{\f{S_i S_j}{4}} \left( \cos(\theta_i +\nu_{ij} \theta_j)+\nu_{ij} \right) \left( a_i a_j +a^\dagger_i a^\dagger_j \right) \nonumber \\
&   & + \sum_i B_i \sin \theta_i a^\dagger_i a_i
\eea
which can be written in the form
\be \label{quadHamiltonian}
H_2 =  \sum_{ij} \left( a^\dagger_i A_{ij} a_j + a_i A^*_{ij} a^\dagger_j +
		      a^\dagger_i B_{ij} a^\dagger_j + a_i B^*_{ij} a_j \right) + G
\ee
where the constants are
\be
G = - \sum_i \left( \f{B_i}{2} \sin \theta_i + \sum_{j=e_i} \f{J_{ij}}{2} \nu_{ij} \cos(\theta_i+\nu_{ij} \theta_j) S_j \right),
\ee
\bea
A_{ij} & = & J_{ij} \f{\sqrt{S_i S_j}}{4} \left( \cos(\theta_i +\nu_{ij} \theta_j) - \nu_{ij} \right) \delta_{<ij>} \\
&   & + \left( \f{B_i}{2} \sin \theta_i + \sum_{k=e_i} \f{J_{ik}}{2} \nu_{ik} S_k \cos(\theta_i+\nu_{ik} \theta_j) \right) \delta_{ij} \nonumber 
\eea
and
\be
B_{ij} =  J_{ij} \f{\sqrt{S_i S_j}}{4} \left( \cos(\theta_i +\nu_{ij} \theta_j) + \nu_{ij} \right) \delta_{<ij>}
\ee
where $\delta_{<ij>}$ is 1 when $i$ and $j$ are nearest neighbors and zero otherwise.

\begin{figure}
\includegraphics[clip,width=4cm]{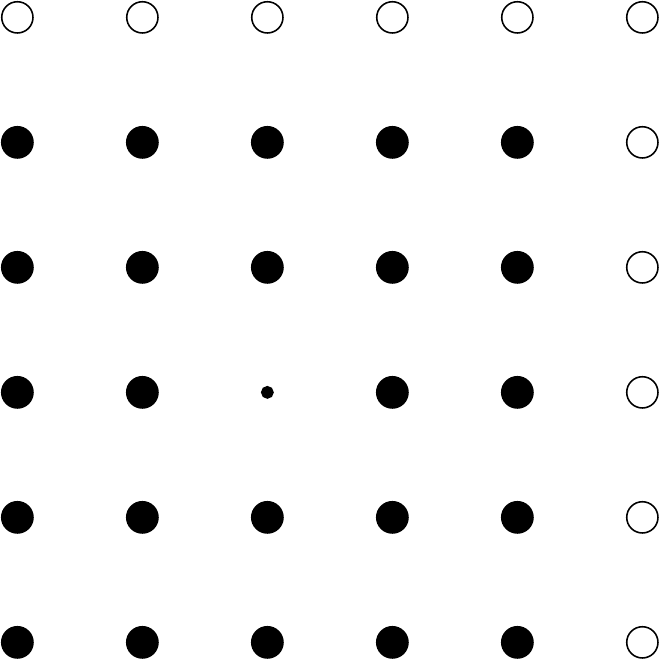} 
\caption{ 
Geometry of a $6 \times 6$-lattice. The open circles mark sites where the boundary condition is imposed. The filled circles are sites where the angles are being calculated. The small circle is the impurity site. Periodic boundary conditions are used.
\label{pbc}} 
\end{figure}

In order to numerically diagonalize Eq.~(\ref{quadHamiltonian}) we will first find the numerical values of the $\theta_i$'s by solving Eq.~(\ref{allminimal}). This is achieved by the relaxation method where the boundary condition is specified as $\sin \theta_{\rm boundary}= B/2SZJ$ and an initial guess for the angles on other sites is made as indicated in Fig.~\ref{pbc}. Then the lattice is traversed site by site and new angles are computed using Eq.~(\ref{allminimal}). This step is repeated until convergence. It is known that this procedure converges slowly. However for typical lattice sizes ($28 \times 28$) used here this is not an issue of practical importance. Having determined the angles numerically we proceed to diagonalize the quadratic Hamiltonian.

We begin by forming the $2N$ column vector
$\avector = ( a_1, a_2, \ldots, a_N,a^\dagger_1,a^\dagger_2,\ldots,a^\dagger_N)^T$ where we have numbered the lattice sites in a consecutive fashion from $1$ through $N$. 
The components of $\avector$ obey the following commutation relation $\left[ \avector_i,\avector^\dagger_j \right]= \etamatrix_{ij}$ where $\etamatrix= \left( \begin{smallmatrix} 1_{N\times N}& 0 \\ 0 & -1_{N\times N} \end{smallmatrix} \right)$. With this notation
the quadratic Hamiltonian takes the form
\be \label{diagHamiltonian}
H = \avector^\dagger \dmatrix \avector 
\ee
where $\dmatrix$ is the $2N \times 2N$-matrix with entries from the quadratic part of the Hamiltonian
\be 
\mathfrak{D} = \begin{pmatrix} A & B \\ B^* & A^* \end{pmatrix}.
\ee
We seek a $2N \times 2N$ Bogoliubov transformation matrix $\tmatrix$ that transforms $\avector$ into new bosonic operators $\bvector$: $\avector = \tmatrix \bvector$.
In order for the entries of $\bvector$ to obey bosonic commutation rules the matrix $\tmatrix$ must obey
\be \label{commutationcondition}
\etamatrix = \tmatrix \etamatrix \tmatrix^\dagger.
\ee
Inserting $\avector = \tmatrix \bvector$ into the Hamiltonian (\ref{diagHamiltonian}) we seek a $\tmatrix$ that fulfills the commutation condition Eq.~(\ref{commutationcondition}) and that makes $\tmatrix^\dagger \dmatrix \tmatrix = \ematrix$ where $\ematrix$ is diagonal. However it is not always possible to find such a diagonal matrix. When the Hamiltonian contains zero modes associated with a continuous spectrum one will never be able to write the free particle operator $p^2$ as a $b^\dagger b$ term alone. However such a term can always be written as $b^\dagger b + b b^\dagger -b b - b^\dagger b^\dagger$ with the proper rescaling of operators. Thus we will seek a matrix $\ematrix$ that is almost diagonal in the sense that for massive modes it has only entries along the diagonal while the continuous parts of the spectrum is represented by 1s or -1s in appropriate places. More specifically we are seeking a matrix $\tmatrix$ that makes $\tmatrix^\dagger \dmatrix \tmatrix$ into a $2N \times 2N$-matrix $\ematrix$ of the form
\be 
\ematrix =
\begin{pmatrix} 
 E_e &             &     &     &             &     \\
     & 0_{\bar{z}} &     &     & 0_{\bar{z}} &     \\
     &             & I_z &     &             & J_z \\
     &             &     & E_e &             &     \\
     & 0_{\bar{z}} &     &     & 0_{\bar{z}} &     \\
     &             & J_z &     &             & I_z
\end{pmatrix}
\ee
where $E_e$ is a diagonal $e \times e$ matrix of positive energies which represents the discrete harmonic oscillator energies associated with $e$ gapped modes. Here $0_{\bar{z}}$ is a $\bar{z} \times \bar{z}$-matrix of zeros that represents $\bar{z}$ proper zero modes where the harmonic oscillator energy is zero,
$I_z$ and $J_z$ are describing the $z$ improper zero modes associated with a continuous free-particle spectrum, 
$I_z$ is a $z \times z$-diagonal unit matrix, and   $J_z$ is a $z \times z$ diagonal matrix with diagonal entries either $+1$ or $-1$. The sign distinguishes between operators of the type $x^2$ and $p^2$. Empty entries indicate zeros. 
The procedure of finding such a $\tmatrix$ is outlined in details in Ref.~\onlinecite{Colpa1}. We have implemented this on a computer and find that the procedure works very well. 

In the absence of linear terms the magnetization is given to quadratic order by
\be
\langle S^z_i \rangle = \sin \theta_i \left( S_i - \langle a^\dagger_i a_i \rangle \right).
\ee
The value of $\sin \theta_i$ is known from the minimization of the classical term, and $\langle a^\dagger_i a_i \rangle$ can be obtained from the transformation matrix $\tmatrix$. Without loss of generality the matrix $\tmatrix$ can be written
\be
\tmatrix = \begin{pmatrix} U & V^* \\ V & U^* \end{pmatrix}
\ee
where $U$ and $V$ are $N \times N$-matrices. Then the expectation value $\langle a^\dagger_i a_i \rangle$ is
\bea
\langle a^\dagger_i a_i \rangle & = & \sum_{jk} \left( U^*_{ij} U_{ik} \langle b^\dagger_j b_k \rangle
+V_{ij} V^*_{ik} \langle b_j b^\dagger_k \rangle \right. \nonumber \\
&  & \left. +U^*_{ij} V^*_{ik} \langle b^\dagger_j b^\dagger_k \rangle
+V_{ij} U_{ik} \langle b_j b_k \rangle \right).
\eea
We will compute the expectation value in a state with low energy. For massive modes we pick the ground state to be the vacuum state and then only the second term contributes $\langle b_j b^\dagger_k \rangle = \delta_{jk}$.
The situation is not so simple for the improper zero modes.
An improper zero mode $b^\dagger b + b b^\dagger \pm  b b \pm b^\dagger b^\dagger$ can be written as the momentum squared operator $2p^2$ (the minus sign) or the position squared operator $2x^2$ (the plus sign) using $b= \f{1}{\sqrt{2}} \left( x +ip \right)$ and $b^\dagger= \f{1}{\sqrt{2}} ( x-ip)$. Thus it is clear that its spectrum is continuous.  

For each improper zero mode we choose instead to compute the expectation value in a Gaussian state\cite{PWA} characterized by a width $w$. Specifically
\be
\psi(x) = \left( \f{1}{\pi w^2} \right)^{1/4} e^{-\f{1}{2}(x/w)^2}.
\ee
In this state the expectations values of the energies are
\bea
\langle p^2 \rangle & = & \f{1}{2} w^{-2} \\
\langle x^2 \rangle & = & \f{1}{2} w^2 
\eea
while the expectation values of the operators needed in $\langle a^\dagger_i a_i \rangle$ are
\bea
\langle b^\dagger b \rangle & = & \left( w^2+w^{-2} -2 \right)/4    \\
\langle b b^\dagger \rangle & = & \left( w^2+w^{-2} +2 \right)/4    \\
\langle b^\dagger b^\dagger \rangle & = & \langle b b \rangle = \left( w^2 - w^{-2} \right)/4  
\eea
Using this the expectation value $\langle a^\dagger_i a_i \rangle$ takes the form
\bea
\langle a^\dagger_i a_i \rangle & = & \sum_{j \in e} |V_{ij}|^2 
 + \sum_{j \in z} \f{1}{4} \left( w^2_j |U^*_{ij}+ V_{ij}|^2 \right. \nonumber \\ 
 &  & \left. +\f{1}{w^2_j} |U^*_{ij}- V_{ij}|^2 -2\left( |U_{ij}|^2 - |V_{ij}|^2\right) \right). 
\eea
We will refer to the last sum in the above as the zero mode(s) contribution, and we have allowed for a separate width $w_j$ for each improper zero mode. We will choose values of $w_j$ so that the total energy of the improper zero modes is equal to that of the lowest finite energy mode. This choice is made to avoid divergences and at the same time still justify calling them zero energy modes. In our case, in the presence of a magnetic field,  there is only one improper zero mode, and it turns out that the precise value of the $w$ is not important quantitatively for the z-axis magnetization. In all cases we have looked at here, the zero mode contribution is negligible and we might as well neglect it completely. This is in contrast to the one dimensional case where the zero modes dominate and are responsible for the divergences of spin-wave theory in the infinite volume limit.

\begin{figure}[h]
\includegraphics[clip,width=8cm]{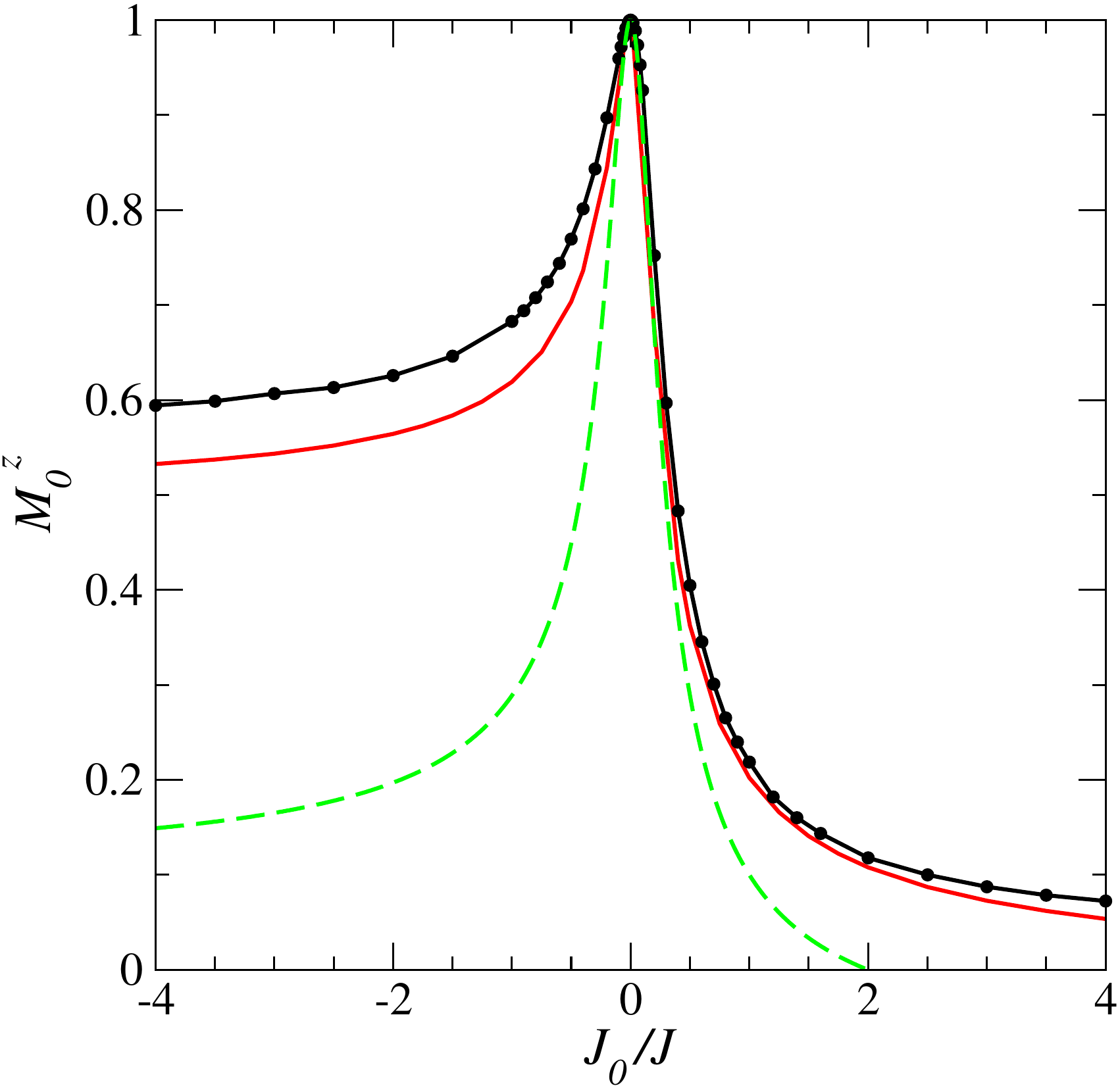} 
\caption{(color online) Magnetization at the impurity site for a spin-1 impurity coupled to a bulk spin-1/2 antiferromagnet by a coupling $J_0$. The filled black circles are results from Quantum Monte Carlo simulations. The dashed line (green) is the classical result coming from Eq.~(\ref{classicalmag}), and the solid line (red) is the numerical spin wave result.  
\label{impmagnS1}} 
\end{figure}

The results from this numerical diagonalization on a $28 \times 28$-lattice is 
shown in Fig.~\ref{impmagnS12} for $S_0=1/2$ alongside the classical result and results from QMC simulations for the square lattice at a fixed value of the magnetic field $B/J=0.4$. Fig.~\ref{impmagnS1} is similar but for $S_0=1$. One can see that the numerical diagonalization procedure compares much more favorably to the QMC data than the classical result does. Especially for antiferromagnetic $J_0$ the agreement is very good. For large ferromagnetic $J_0$ the agreement is worse 
which we believe is related to the truncation of the Hamiltonian at quadratic order in boson operators. The main feature of the curves is a maximum at $J_0$ which reflects the trivial fact that an uncoupled (isolated) impurity will point along the magnetic field. In fact the impurity spin will point along the field for most couplings except very large antiferromagnetic $J_0$ for $S_0=1/2$.

\begin{figure}[h]
\includegraphics[clip,width=8cm]{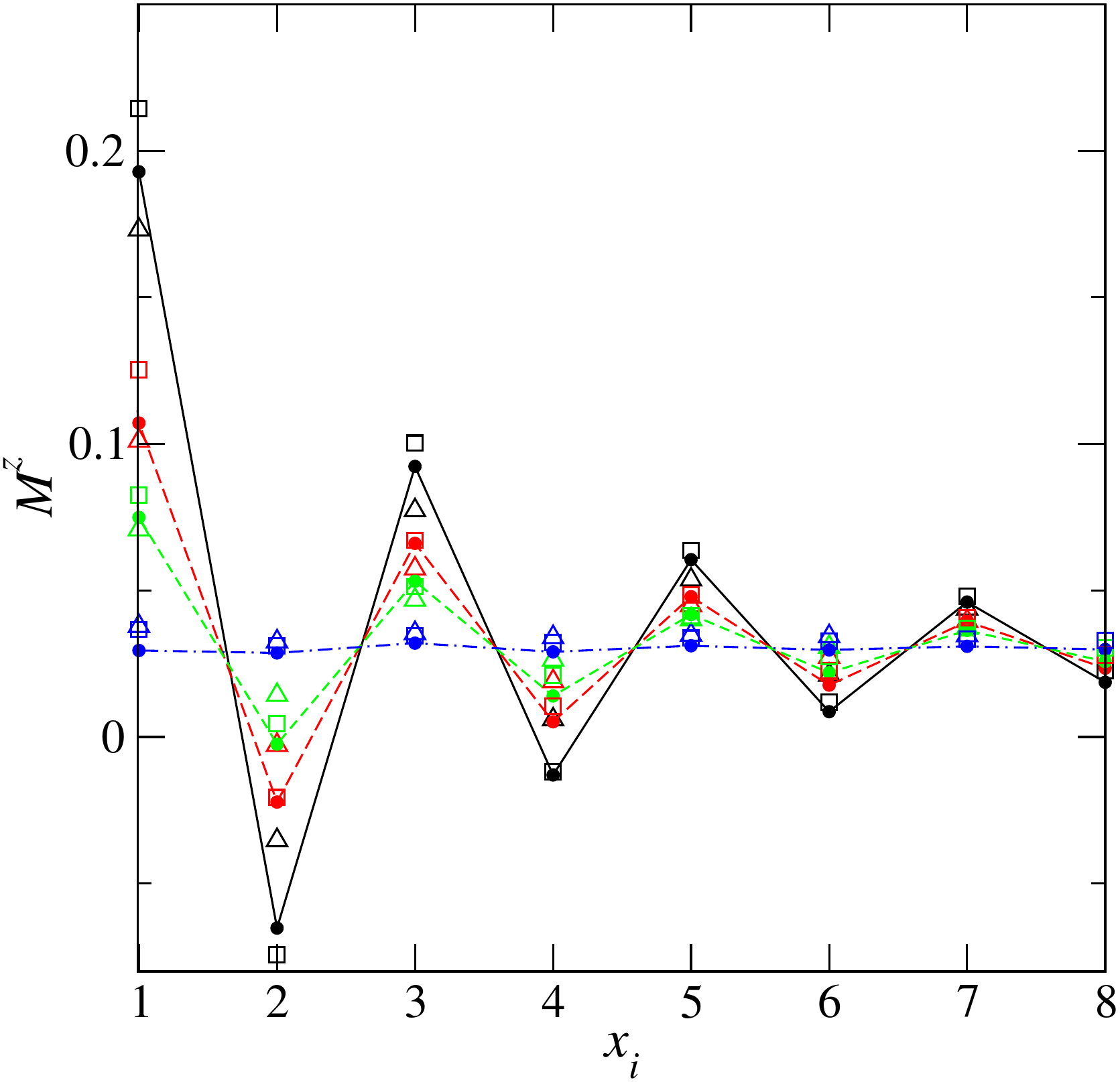} 
\caption{(color online) Magnetization as a function of horizontal distance $x_i$ from the impurity site as calculated by QMC (solid circles), numerical spin waves (triangles) and the anaytic spin wave theory (squares). $S_0=1/2$, $S=1/2$ and $B=B_0=0.4J$. The colors are for different values of $J_0/J=$; $-2$(solid, black), $0$(long dashed, red), $0.1$ (dot-dashed, green) and $0.5$ (dot-dashed, blue). QMC error bars are smaller than the size of the solid circles, and both the QMC and the numerical spin wave calculations are carried out on a $28 \times 28$- lattice.  
\label{horizontalS12}} 
\end{figure}

\begin{figure}[h]
\includegraphics[clip,width=8cm]{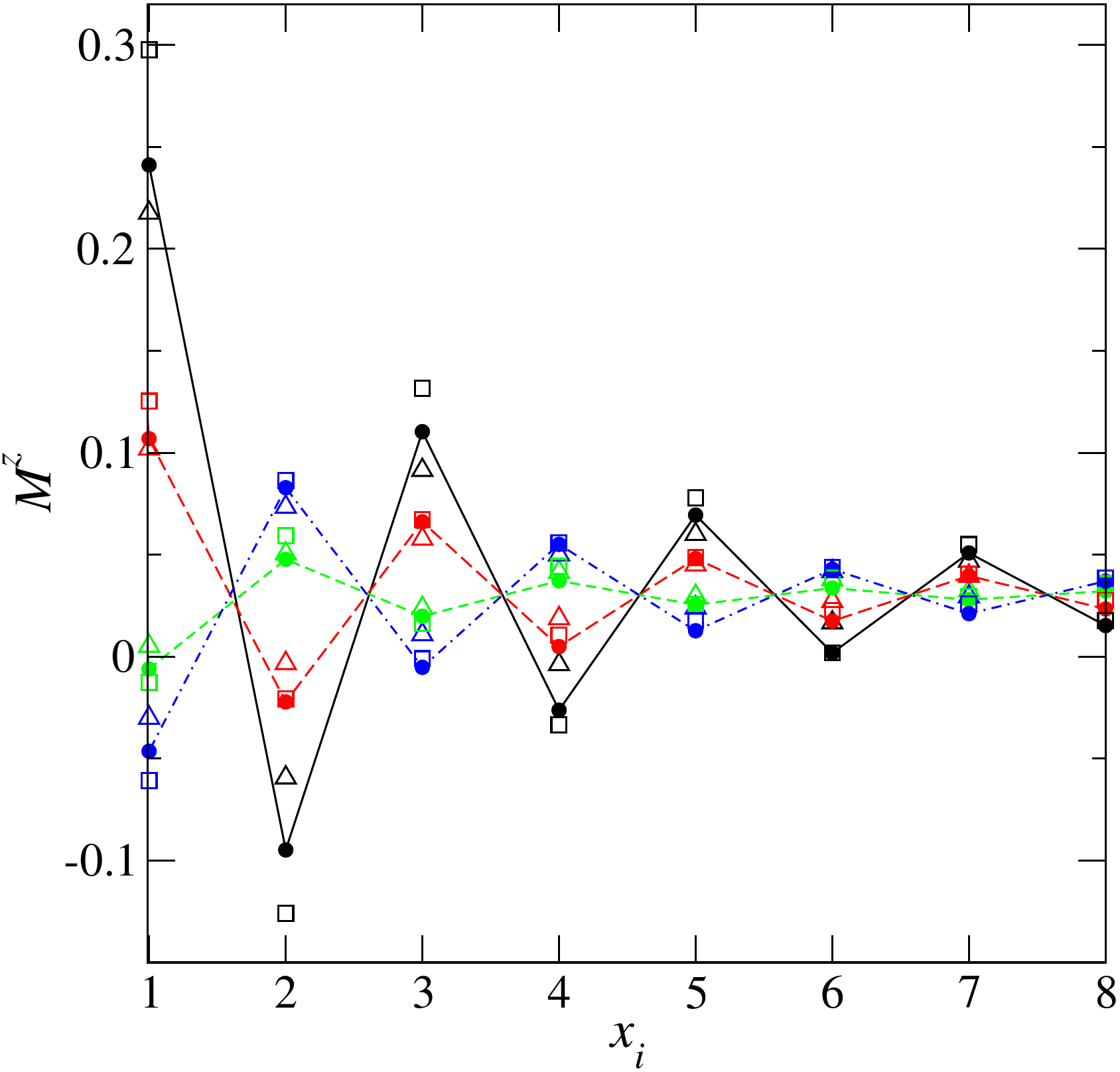} 
\caption{(color online) Magnetization as a function of horizontal distance $x_i$ from the impurity site as calculated by QMC (solid circles), numerical spin waves (triangles) and the anaytic spin wave theory (squares). $S_0=1$, $S=1/2$ and $B=B_0=0.4J$. The colors are for different values of $J_0/J=$ $-1$(solid, black), $0$(long dashed, red), $0.2$ (dashed, green) and $1$ (dot-dashed, blue). QMC error bars are smaller than the size of the solid circles, and both the QMC and the numerical spin wave calculations are carried out on a $28 \times 28$- lattice.  
\label{horizontalS1}} 
\end{figure}

For sites in the neighborhood of the impurity we can also compare the analytic and the numerical spin wave calculation to the QMC results. In Fig.~\ref{horizontalS12} we show the magnetization for an $S_0=1/2$ impurity at different positions $(x_i,y_i=0)$ close to the impurity. The different lines are for the various values of the impurity coupling $J_0$ and the different symbols indicate the method used. In comparing the methods we see that  the analytic result lies reasonably close to the QMC data except for the nearest neighbor point where the numerical spin wave calculation give a better approximation to the QMC data. For a fixed value of $J_0$ one can see that the magnetization exhibits a predominantly alternating pattern with a  magnitude that is largest for ferromagnetic couplings $J_0<0$ as predicted in Fig.~\ref{C}. As the ferromagnetic coupling $J_0$ becomes smaller the magnetization of the impurity spin increases, Fig.~\ref{impmagnS12}, while the surrounding pattern is not much affected. On the antiferromagnetic side, $J_0 >0$, the magnetization of the impurity spin 
decreases accompanied also by a decrease in the amplitude of the magnetization oscillation away from the impurity. At $J_0=J$ the oscillation pattern vanishes completely. For strong antiferromagnetic couplings $J_0 > J$ there is almost no induced magnetization on the sites surrounding the impurity, but the magnetization of the impurity spin 
becomes smaller than the average magnetization and can even become negative for strong enough $J_0$.

For the $S_0=1$ impurity the magnetization pattern around the impurity is shown in Fig.~\ref{horizontalS1}. Again the oscillations are large for ferromagnetic $J_0$. As $J_0 \to 0$ the magnetization of the impurity spin increases while the oscillating pattern around it decreases. Then as $J_0$ becomes antiferromagnetic the magnetization oscillations increase again, but now the sublattice pattern has changed to the pattern in Fig.~\ref{spins} b), consistent with the fact that $C$ changes sign in Fig.~\ref{C}. The amplitude of the oscillations saturates as $J_0$ becomes even stronger.

\section{Discussion}
We have presented results for the magnetization around a general impurity in a Heisenberg spin-S antiferromagnet in a magnetic field. Away from the impurity we find that the induced magnetization is dominantly a staggered magnetization in the field direction. We have calculated this alternating magnetization, and our results are in reasonable agreement with extensive QMC simulations that we have also carried out. 
One important feature of the spin wave result 
is that the parameters of the impurity model only affect the overall prefactor $C$
of the magnetization while the scale and shape of the decay are universal and 
only reflect the properties of the host magnet and the applied field. 
We have analyzed how the prefactor $C$ depends on impurity properties and  
found that the effect on the alternating magnetization is largest for ferromagnetically coupled  impurities
and  generally increases with magnetic field. 
In order to calculate the magnetization at the impurity site 
we have described in detail how to diagonalize the quadratic spin wave Hamiltonian numerically. This approach agrees well with the QMC calculations and we have outlined how the magnetization of the impurity spin depends on the coupling strength of the impurity to its neighbors. 

In summary the results can be used to predict the detailed local magnetization pattern around
general magnetic and non-magnetic impurities in isotropic antiferromagnets, 
e.g. from doping Zn, Co and Ni in 
copper-oxide antiferromagnets.  In most real materials the effects from crystal fields and 
other anisotropies are also important, but our calculations provide
the first step, before other possible terms in the Hamiltonian are taken into account.

\begin{acknowledgments}
The QMC calculations were carried out on CPUs provided by the Notur project.
Financial support by
the DFG via the research initiative SFB-TR49 and the 
Graduate School of Excellence MAINZ/MATCOR is gratefully acknowledged.
\end{acknowledgments}

\appendix*
\section{Sum}
The sum 
\be
I  = \f{1}{N} \sum_{\vec{k}} \f{\gamma_{\vec{k}}}{1 + g \gamma_{\vec{k}}} e^{i \vec{k} \cdot \vec{r}}
\ee
for $\vec{r} \neq 0$ can be written
\be
I  =  \f{1}{gN} \sum_{\vec{k}} \f{1+g\gamma_{\vec{k}}-1}{1 + g \gamma_{\vec{k}}} e^{i \vec{k} \cdot \vec{r}}
= -\f{1}{gN} \sum_{\vec{k}} \f{1}{1 + g \gamma_{\vec{k}}} e^{i \vec{k} \cdot \vec{r}}.
\ee
This sum can be calculated by expanding the denominator about the antiferromagnetic point 
 $\vec{Q}=(\pi,\pi,\ldots)$. 
Shifting the $\vec{k}$ summation $\vec{k} \to \vec{k}+\vec{Q}$ 
and expanding the denominator to order $\vec{k}^2$ we get
\be
I \approx - \f{e^{i \vec{Q} \cdot \vec{r}}}{g}\f{1}{N} \sum_{\vec{k}}
   \f{e^{i \vec{k} \cdot \vec{r}}}{1-g+ g \vec{k}^2/Z}
\ee
where $Z$ is the coordination number of the lattice. This can also be written
\be
I \approx - \f{Z d^2 e^{i \vec{Q} \cdot \vec{r}}}{g^2}\f{1}{N} \sum_{\vec{k}}
   \f{e^{i \vec{k} \cdot \vec{r}}}{1+ d^2 \vec{k}^2}
\ee
where $d=\sqrt{\f{g}{Z(1-g)}}$.
The sum is calculated by transforming it into an integral and using polar coordinates
\be
\f{1}{N} \sum_{\vec{k}} \f{e^{i \vec{k} \cdot \vec{r}}}{1+d^2 \vec{k}^2}
= \f{1}{2\pi d^2} \left\{ 
\begin{matrix} 
	  K_0(r/d),  & \; D=2 \\
~& ~\\
	  e^{-r/d}/(2r),& \; D=3
\end{matrix} \right.
\ee
where $K_0$ is the zeroth order modified Bessel function of the second kind.
Putting this together we get
\be
I \approx - \f{Z e^{i \vec{Q} \cdot \vec{r}}}{2\pi g^2}
\left\{ 
\begin{matrix} 
  K_0(r/d),  & \; D=2 \\
~& ~ \\
  e^{-r/d}/(2r), & \; D=3,
\end{matrix} \right.
\ee 
where  $e^{i \vec{Q} \cdot \vec{r}} = (-1)^{x_i+y_i+z_i}$.

\end{document}